\documentclass[structabstract]{aa}  

\usepackage{graphicx}
\usepackage{txfonts}
\usepackage{natbib}
\usepackage{lscape}
\usepackage{longtable,lscape}
\begin{document}
   \title{Identification of red high proper-motion objects in Tycho-2
     and 2MASS catalogues using Virtual Observatory tools}

   \titlerunning{Red high proper-motion objects in Tycho-2 and 2MASS}

   \author{F. M. Jim\'enez-Esteban\inst{1,2,3}
     \and
     J. A. Caballero\inst{1}
     \and
     R. Dorda\inst{4,5}
     \and 
     P. A. Miles-P\'{a}ez\inst{4,6}
     \and
     E. Solano\inst{1,2}
}
   \institute{
     Centro de Astrobiolog\'{\i}a (INTA-CSIC), Departamento de Astrof\'{\i}sica,
     PO Box 78, E-28691, Villanueva de la Ca\~nada, Madrid, Spain\\ 
     \email{fran.jimenez-esteban@cab.inta-csic.es}
     \and
     Spanish Virtual Observatory
     \and
     Saint Louis University, Madrid Campus, Division of Science and Engineering,
     Avenida~del~Valle 34, E-28003, Madrid, Spain
     \and 
     Departamento de Astrof\'{\i}sica y Ciencias de la Atm\'{o}sfera, Facultad de
     Ciencias F\'{\i}sicas, Avenida Complutense s/n, E-28040, Madrid, Spain 
     \and 
     Departamento de F\'{\i}sica, Ingenier\'{\i}a de Sistemas y
     Teor\'{\i}a de la Se\~nal, Escuela Polit\'{e}cnica Superior, University of
     Alicante, Apdo.~99, E-03080, Alicante, Spain
     \and 
     Instituto de Astrof\'{\i}sica de Canarias, E-38200, La Laguna, Tenerife, Spain
}
   \date{Received 01 Nov 2011 / Accepted .. Jan 2012}

 
  \abstract
     {}
    {With available Virtual Observatory tools, we looked for new M
     dwarfs in the solar neighbourhood and M giants with high
     tangential velocities.}
    {From an all-sky cross-match between the optical Tycho-2 and
      the near-infrared 2MASS catalogues, we selected objects with proper
      motions $\mu$\,$>$\,50\,mas\,yr$^{-1}$ and very red $V_T-K_{\rm
        s}$ colours. For the most interesting targets, we collected
      multi-wavelength photometry, constructed spectral energy
      distributions, estimated effective temperatures and surface
      gravities from fits to atmospheric models, performed time-series
      analysis of ASAS $V$-band light curves, and assigned spectral
      types from low-resolution spectroscopy obtained with CAFOS at
      the 2.2\,m Calar Alto telescope.}
      {We got a sample of 59 bright red high proper-motion objects,
        including fifty red giants, four red dwarfs, and five objects
        reported in this work for the first time. The five new stars
        have magnitudes $V_T$\,$\approx$\,10.8--11.3\,mag, reduced
        proper motions midway between known dwarfs and giants,
        near-infrared colours typical of giants, and effective
        temperatures $T_{\rm eff}$\,$\approx$\,2900--3400\,K. From our
        time-series analysis, we discovered a long secondary period in
        Ruber\,4 and an extremely long primary period in
        Ruber\,6. With the CAFOS spectra, we confirmed the red giant
        nature of Ruber\,7 and 8, the last of which seems to be one of
        the brightest metal-poor M giants ever identified.}
     {}
 
 
    \maketitle

%

\section{Introduction}
\label{introduction}

For the Virtual Observatory (VO), there are seductive mottoes
(e.g. {\em the Universe at your fingertips}; NVO\footnote{\tt
  http://www.us-vo.org}) and bombastic definitions (e.g. {\em an
  international astronomical community-based initiative [that] aims to
  allow global electronic access to the available astronomical data
  archives of space and ground-based observatories and other sky
  survey databases [and] to enable data analysis techniques through a
  coordinating entity that will provide common standards, wide-network
  bandwidth, and state-of-the-art analysis tools};
EURO-VO\footnote{\tt http://www.euro-vo.org}). One of the most
tangible results of the VO endeavour is a suite of analysis tools (the
so-called ``VO tools''), which are used more and more frequently by
the astronomical community, as demonstrated by the growing number of
VO papers\footnote{\tt http://www.euro-vo.org/pub/fc/papers.html}
(see, for instance, \citealt{Caballero09} for a good example of this).

While there is a current trend to discovering the ``faintest, coolest,
smallest'' objects in our Galaxy
(e.g. \citealt{Delorme08a,Kirkpatrick11}), there is still a lot to do
with relatively bright unknown objects ($V \gtrsim$ 11\,mag --
i.e. visible through small amateur telescopes). With
\cite{Caballero08} and \cite{Jimenez-Esteban11}, we started a project
devoted to identifying bright {\em blue} objects in the Tycho-2
catalogue interesting for other follow-up studies. In particular, our
surveys have provided some of the brightest hot subdwarfs ever found,
which enormously facilitates forthcoming astroseismologic,
spectroscopic, and multiplicity analyses \citep{Oreiro11,Vennes11a}.

Here we go on examining high proper-motion objects in Tycho-2 and
2MASS catalogues using VO tools, this time to the other side of the
colour-magnitude diagram, looking for bright {\em red} objects. In
the absence of discs or dust envelopes, red colours are synonymous
with low effective temperatures ($T_{\rm eff} \sim$ 3000\,K for the
coolest stars in Tycho-2). A priori, there are not many alternatives
for classifying such bright late-type objects with high proper-motion:
they can be either nearby M dwarfs or red giants with high tangential
velocities.

Finding a single uncatalogued Tycho-2 dwarf later than M3V would
justify the whole current survey. With $V \sim$ 11\,mag, a star like
that would have $J \lesssim$ 7\,mag and inevitably be a golden target
for the next high-resolution, near-infrared (NIR), radial-velocity
surveys seeking Earth-like planets
\citep{Reiners10,Mahadevan10,Quirrenbach10}. Such a finding was also
one of the aims of \cite{Lepine11}, who presents an all-sky catalogue
of M dwarf stars with apparent infrared magnitudes $J <$ 10.0\,mag.
In a sense, the \cite{Lepine11} catalogue and ours are complementary.

The other alternative is a late-type giant (spectral class M, C, or
S). Since giants are intrinsically luminous, they are commonly located
at long heliocentric distances. However, if they have appreciable
proper motions, of over a few tens of milliarcseconds per year, their
tangential velocities must be very high. As a result, they may belong
to the Galactic thick disc or even the halo
\citep{Carney86,Rocha-Pinto04,Famaey05}. Our work is expected to
catalogue all the brightest, reddest giants over the whole sky, which
are excellent targets for investigating the dynamics and metallicity
structure of the Milky Way \citep{McWilliam94,Lambert86,Beers02} and
their interiors through pulsation analyses
\citep{Wood77,Bergeat02,Kiss03}. Besides this, we also expect to
discover rare examples of bright red giants.


\section{Analysis and results}
\label{analysisandresults}

\subsection{Target selection}
\label{targetselection}

\begin{figure*}
\centering
\includegraphics[width=0.99\textwidth]{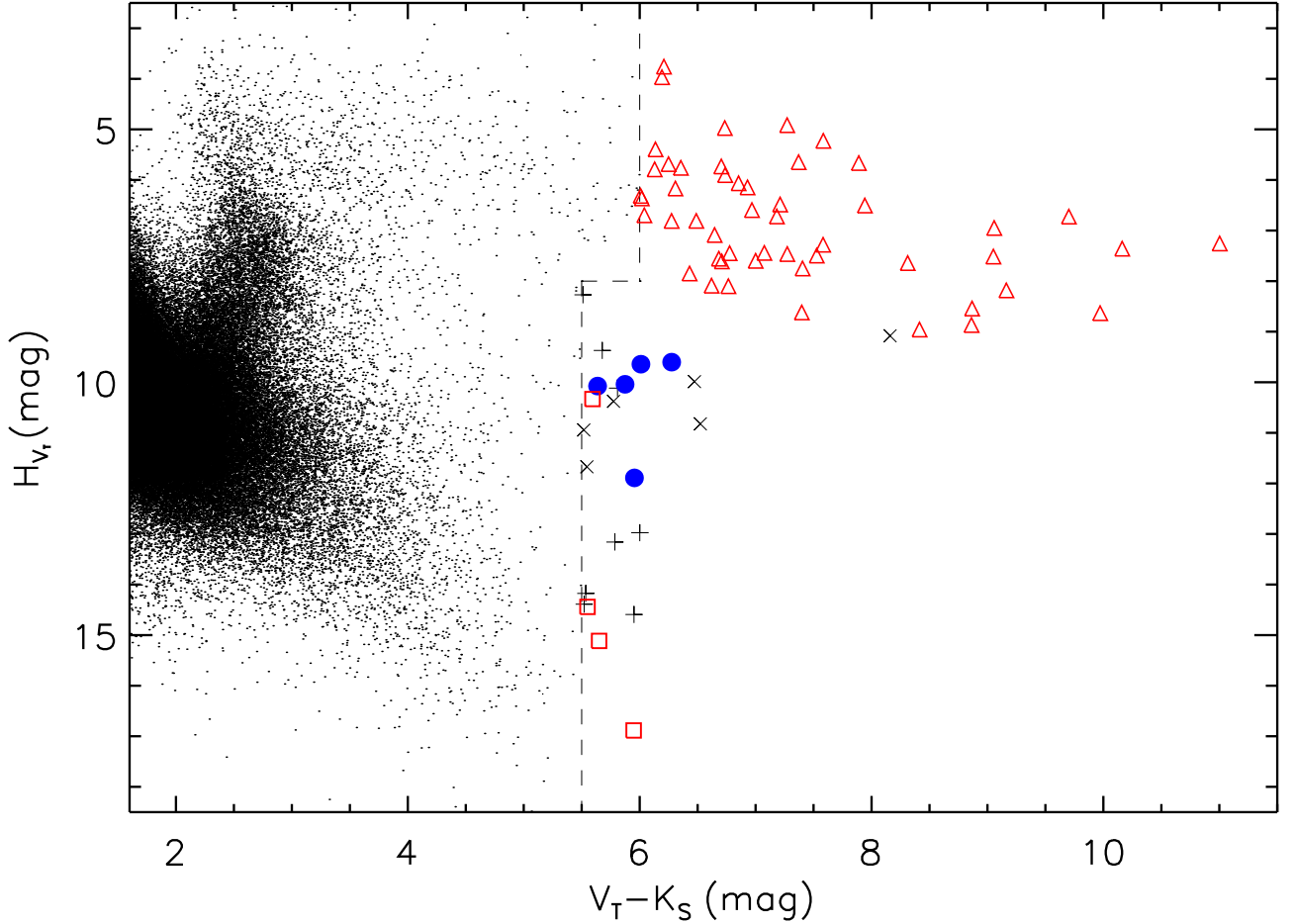}
\caption{Tycho-2/2MASS reduced proper motion diagram ($H_{V_T}$ versus
  $V_T-K_S$, where $H_{V_T} = V_T + 5 \log{\mu} + 5$). Objects
  selected for follow-up are redwards of the dashed line. Previously
  unreported red high proper-motion objects are depicted with [blue]
  filled circles, already-known objects with [red] open symbols
  (giants with triangles, dwarfs with squares), binary systems
  unresolved by 2MASS with [black] crosses, and objects with wrong
  proper motion with [black] sails. The remaining objects bluewards of
  the dashed line are marked with small dots. The dwarf located
  halfway between giants (to the top) and the remaining dwarfs (to the
  bottom) is BD--21~1074~BC. Compare this diagram with the one in
  Fig. 2 of \cite{Jimenez-Esteban11}.}
\label{rpm}
\end{figure*}

We used the sample of 155,384 sources with proper motions
$\mu$\,$>$\,50\,mas\,a$^{-1}$ obtained from the cross-match of the
whole Tycho-2 \citep{Hog00a} and 2MASS \citep{Skrutskie06} catalogues
constructed by \cite{Jimenez-Esteban11}. We selected red high
proper-motion star candidates based on their position on the reduced
proper motion-colour diagram shown in Fig.~\ref{rpm}. Quantitatively,
the applied selection criterion was $V_T-K_{\rm s} >$ 6.0\,mag if
$H_{V_T} \le$ 8\,mag and $V_T-K_{\rm s} >$ 5.5\,mag if $H_{V_T} >$
8\,mag. The abnormally large errors in $K_{\rm s}$ because of
saturation of the brightest objects barely affect the data point
location in the diagram.
The application of this simple selection criterion resulted in 73
objects with redder colours than these values, which is an appropriate
number of sources to be studied individually on a reasonable time
scale.


Next, we carried out a visual inspection of the 73 selected candidates
in a similar way to the one described in \cite{Jimenez-Esteban11}. Six
of them were found to have erroneous proper motions in the Tycho-2
catalogue. As an example, the catalogue gives a proper motion
($\mu_\alpha \cos{\delta}$,~$\mu_\delta$) =
(--14,~--84)\,mas\,a$^{-1}$ for the star TYC~6238--480--1, while we
measured (+0.9$\pm$1.4,~$-12.4\pm$0.6)\,mas\,a$^{-1}$ using six
astrometric epochs covering 43.7 years and the method exposed by
\cite{Caballero10c}. In this case, the Tycho-2 measurement was likely
affected by a visual (unbound) companion at 8\,arcsec to the North,
$\sim$2.8\,mag fainter in the $R$ band.  The five other discarded
stars have even lower actual proper motions.
In addition, we discarded another eight close binary stars, resolved
by Tycho-2 but unresolved by 2MASS, because of their incorrect
resulting colours.

\addtocounter{table}{1} 

\subsection{Preliminary target classification}

Of the 59 remaining unresolved objects, 54 have some kind of published
spectral type information.  We list their basic properties in
Tables~\ref{table.known.giants} and~\ref{table.known.dwarfs}, where we
provide their Tycho-2 identification, coordinates, Tycho-2 $V_T$ and
2MASS $JHK_{\rm s}$ magnitudes, total proper motion $\mu$, parallax
(from \citealt{vanLeeuwen07} in all but two dwarfs), spectral type,
most common name, and at least one of the most representative
references.


Most of the known reddest Tycho-2/2MASS stars with
$\mu$\,$>$\,50\,mas\,a$^{-1}$ are giants. Except for a few cases, the
fifty stars in Table~\ref{table.known.giants} are well known class-III
giants (such as Mira~Cet~AB, L$_2$~Pup, $g$~Cen, and R Lyr, which are
the only stars in the table at $d <$ 100\,pc), semi-regular pulsating,
irregular, or Mira variables. They are located at long parallactic
distances, have strong flux densities in the {\em IRAS} catalogue of
point sources, saturate in 2MASS $JHK_{\rm s}$ bands, and/or already
have spectral type and class determinations (mostly found in the
spectral type compilation by \citealt{Kwok97}) consistent with its
location in the reduced proper motion diagram far from the main
sequence. The only exceptions to the collection of stars with evidence
of ``giantism'' are the poorly investigated stars BD+31~1540 and
HD~150184. The former was already presented as a star with a
remarkable spectrum by \cite{Espin1892} in the 19th century, and it
has the same properties as the rest of the stars in the table. The
latter does not have a SIMBAD entry at all, but was tabulated as an
M3--4 giant star by \cite{Houk75}.

The four other known stars, shown in Table~\ref{table.known.dwarfs},
are M dwarfs in the solar neighbourhood. While two of them have
accurate parallactic distances measured by {\em Hipparcos} (QY~Aur~AB:
$d$ = 6.29$\pm$0.12\,pc; YZ~CMi: $d$ = 5.95$\pm$0.07\,pc;
\citealt{vanLeeuwen07}), the other two only have photometric distances
placing them at 8--10\,pc \citep{Reid95b,Beuzit04}. Three of the four
stars have proper motions larger than any giant in
Table~\ref{table.known.giants}, and one of them up to
1000\,mas\,a$^{-1}$.  The four stars have spectral type determinations
at M3.5\,--\,4.5V, consistent with the observed optical and NIR
colours (in contrast to the giants, none of the dwarfs saturate in the
2MASS $JHK_{\rm s}$ bands).  Interestingly, the four dwarfs are
(candidate) members in multiple systems and/or young moving groups:

\begin{itemize}
\item BD--21~1074~BC is part of a hierarchical triple system,
  WDS~05069--2135, which was discovered by \cite{Donner35}. The
  primary (A, GJ~3331) is a nearby flare M1.5V star at 8.22\,arcsec to
  a tighter system (BC, GJ~3332) of two slightly cooler stars
  separated by only 0.80\,arcsec. Neither Tycho-2 nor 2MASS were able
  to resolve the tight binary, which is 1.15\,mag redder than the
  primary in $V_T-K_{\rm s}$. Based on a significant amount of Li~{\sc
    i} in absorption in the spectra of both A and BC components, X-ray
  emission, and kinematics, \cite{daSilva09} classified the system as
  a member of the very young $\beta$~Pictoris association ($\tau \sim$
  12\,Ma).
\item QY Aur AB is a ``classic'' double in the solar neighbourhood.
  It is a spectroscopic binary with an approximate period of 10.43\,d
  and a relatively high orbital eccentricity of 0.34 with no previous
  report of membership in a young moving group.
\item YZ~CMi, although it has a Washington Double Star catalogue
  entry, seems to be single \citep{Lafreniere07}.  It is an
  emission-line dwarf with megaflares, which has been repeatedly
  investigated (e.g., \citealt{Kunkel69,Kahler82,Kundu88,Benz91}).
  \cite{Montes01} classify the star as a member of the Local
  Association ($\tau \sim$ 10\,--\,150\,Ma).
\item The magnetically active binary star DG~CVn~AB (WDS~13318+2917),
  one of the only three known M dwarf ultra-fast rotators within
  10\,pc \citep{Delfosse98}, was first resolved by \cite{Beuzit04}
  into a pair separated by only 0.17\,arcsec. \cite{Montes01}
  classified the binary as a member of the young disc ($\tau \sim$
  600\,Ma).
\end{itemize}

The remaining five unknown stars, listed in Table~\ref{table.unknown}
and plotted in Fig.~\ref{rpm} with filled (blue) circles, have no {\em
  Hipparcos} entry and have never been reported in the
literature. Actually, one of them has a Bonner Durchmusterung entry
\citep{Schonfeld1886}, but it has gone without being noticed for over
120 years.  For naming the stars, we followed the ``Ruber'' ({\em red}
in Latin) nomenclature introduced by \cite{Caballero08} and followed
by \cite{Caballero09}. Our new Ruber objects go from the fourth to the
eighth in this series.

\begin{figure*}
\centering
\includegraphics[width=0.99\textwidth]{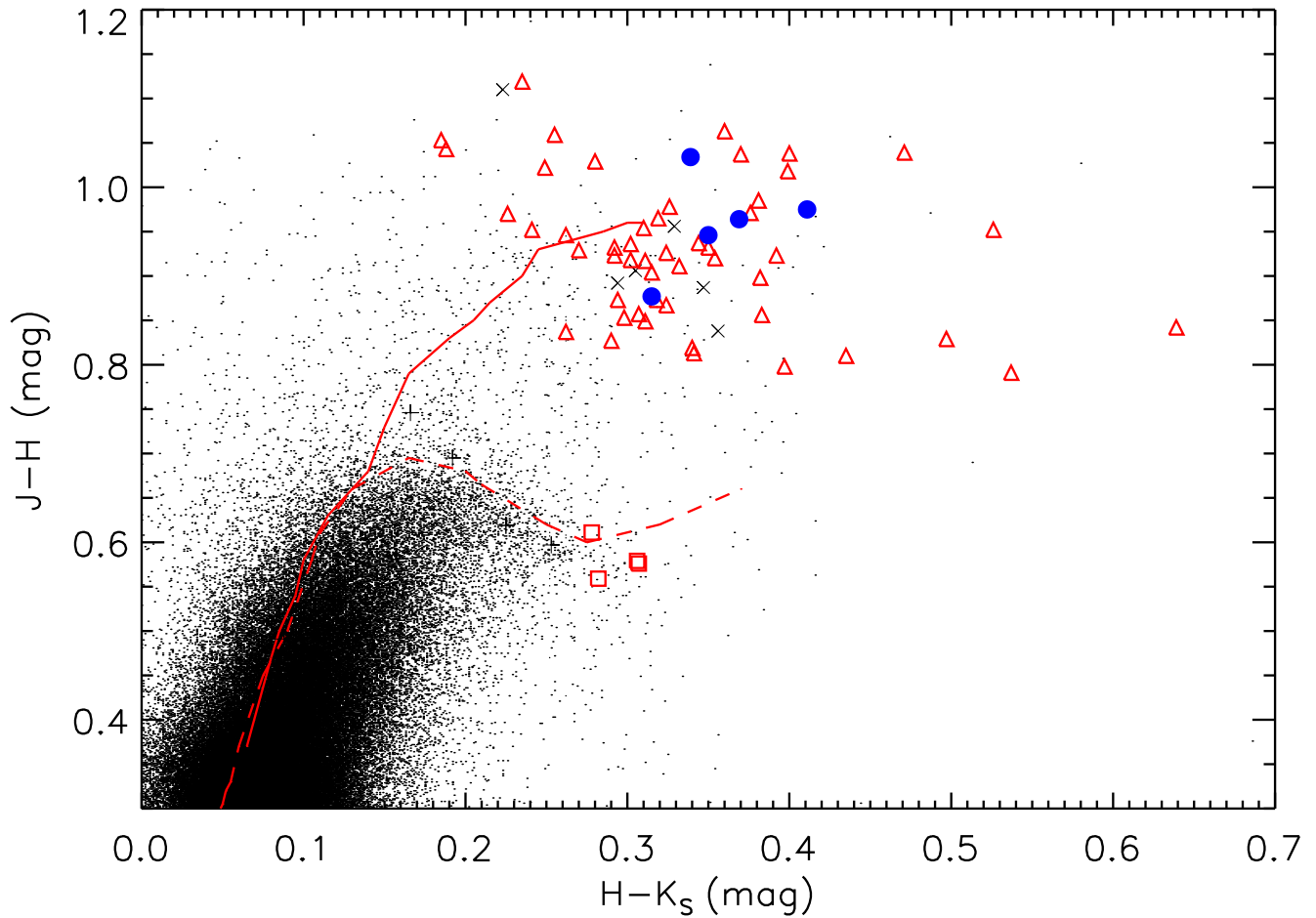}
\caption{Same as Fig.\,\ref{rpm} but for the near-infrared
  colour-colour diagram.  Dashed and continuous lines depict the
  intrinsic colours of M dwarfs and giants, respectively
  \citep{Bessell88}.  The two reddest dwarfs, QY~Aur~AB and YZ~CMi~AB,
  have almost coincident near-infrared colours.}
\label{NIRcc}
\end{figure*}

The five new red high proper-motion objects fall between the giant and
(ordinary) dwarf branches in the reduced proper motion-colour
diagram. Are the Ruber objects unresolved binary dwarfs in the
$\beta$~Pictoris moving group, such as BD--21~1074~BC, or peculiar
underluminous giants? To shed some light on their nature, we first
analysed their near-infrared colours.

Figure\,\ref{NIRcc} shows the location of our sources in an $H-K_{\rm
  s}$\,vs.\,$J-H$ colour-colour diagram, which is an excellent
diagnostic tool for distinguishing between red giants and dwarfs
\citep{Bessell88}.  All fifty M giants in
Table\,\ref{table.known.giants} are clearly segregated from the four M
dwarfs in Table\,\ref{table.known.dwarfs}, which match their expected
intrinsic colours.  The five new Ruber objects are located in the
giant region, indicating that they are probably M giants with very
high tangential velocities.


\subsection{Virtual Observatory analysis}

\subsubsection{Spectral energy distributions}

\begin{figure}
\centering
\includegraphics[width=0.49\textwidth]{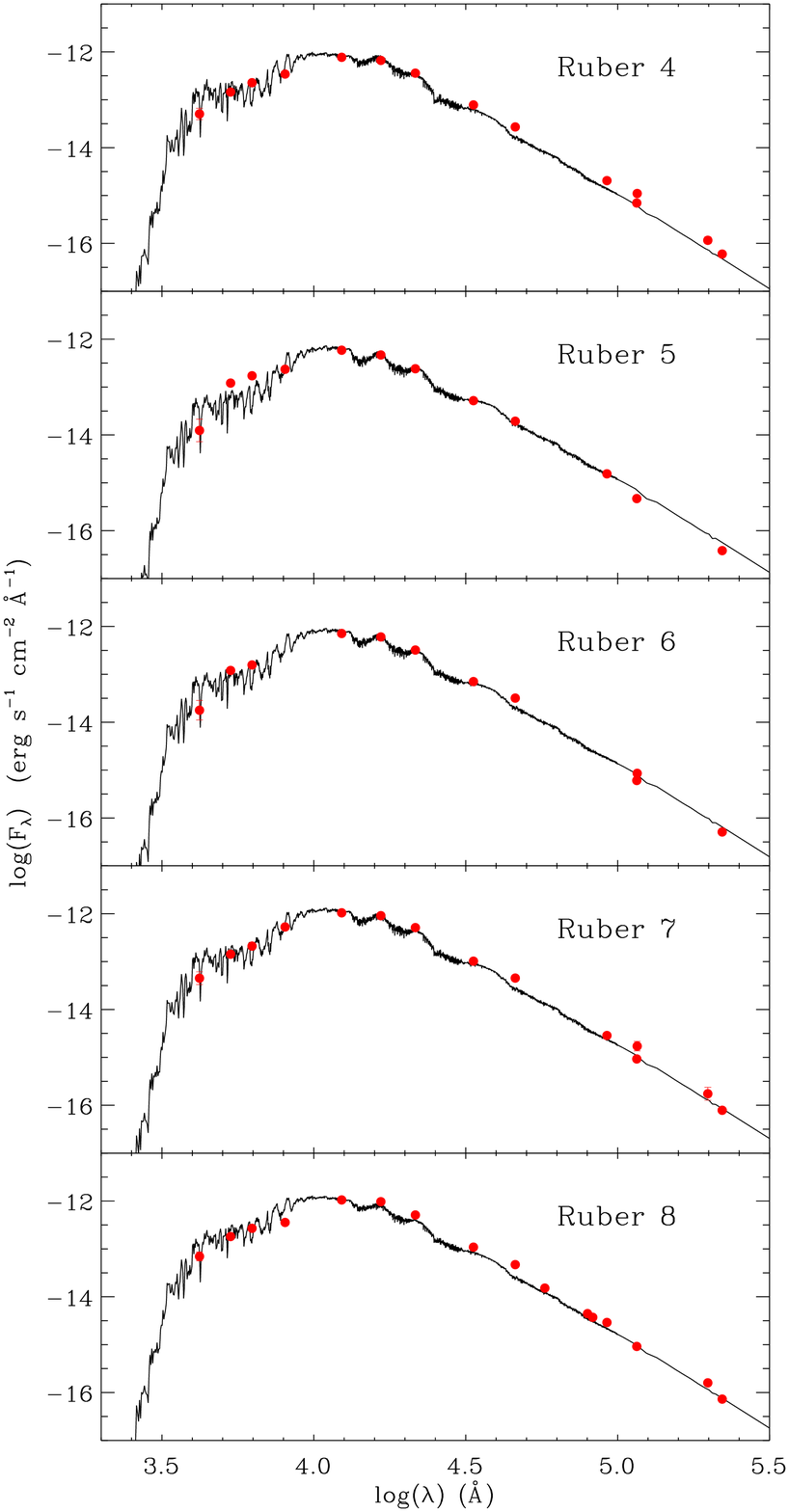}
\caption{Theoretical SED fits of Ruber\,4 to 8, from top to
  bottom. Solid lines represent the NextGen models for solar
  metallicity that best fit the observational data. Filled [red]
  circles indicate the observational photometric data used for the
  fit. Error bars are usually smaller than the size of the circles.}
\label{SEDs}
\end{figure}

Another way to obtain information of the Ruber objects is by analysing
their spectral energy distributions (SEDs). We searched for additional
photometric data of the five Ruber objects using the ``all-VO
Discovery tool'' of Aladin sky atlas \citep{Bonnarel00}. This utility
allows the user to query a large number of photometric catalogues in a
convenient way. Besides Tycho-2 and 2MASS, we collected observational
data from the following astrophotometric catalogues: UCAC3
\citep{Zacharias10}, DENIS \citep{Epchtein97}, {\em WISE}
\citep{Wright10}, GLIMPSE \citep{Churchwell09}, MSX6C \citep{Egan03a},
      {\em AKARI}/IRC \citep{Ishihara10}, and {\em IRAS}
      \citep{Beichman88}. Table~\ref{table.unknown} compiles all this
      information.

We took advantage of another VO tool, VOSA\footnote{\tt
  http://svo.cab.inta-csic.es/theory/vosa} (VO SED Analyzer;
\citealt{Bayo08}), to fit the observed SEDs to theoretical models
available at the VO. VOSA allows the user to query, in an automatic
and transparent way, different collections of theoretical models,
calculate their synthetic photometry, and perform a statistical test
to determine which model reproduces the observed data best.

To fit the SEDs of our Ruber objects, we used the NextGen collection
of stellar atmosphere models \citep{Hauschildt99} and solar
metallicity. The effective temperatures ($T_{\rm eff}$) and surface
gravities ($\log{g}$) obtained with VOSA ranged between 2900 and
3400\,K and between 3.5 and 4.0, respectively (see the bottom of
Table~\ref{table.unknown}). The accuracy in determining these
parameters was set by the model grid size to 100\,K in $T_{\rm eff}$
and 0.5 in $\log{g}$. Estimated temperatures and surface gravities
correspond to late-type M stars with $\log{g}$ midway between dwarfs
and giants. In Figure\,\ref{SEDs} we plot the theoretical fitting of
the observational SEDs.  A VOSA upgrade, with new grids of theoretical
isochrones for dwarf and giant stars with lower $T_{\rm eff}$s than
those provided by the Kurucz models and wide intervals of metallicity,
is currently under development by the Spanish Virtual Observatory team
but, unfortunately, was not available to us at the time of
writing. Nevertheless, the effect of the metallicity on the overall
SEDs {\em of dwarfs} would probably be not strong enough to produce a
noticeable change in the $T_{\rm eff}$ and $\log{g}$ fit values,
especially when fitting broad-band photometry
\citep{Bonfils05,Rojas-Ayala11,Schlaufman10,Woolf09}.


\subsubsection{Photometric variability}

\begin{figure}
\centering
\includegraphics[width=0.49\textwidth]{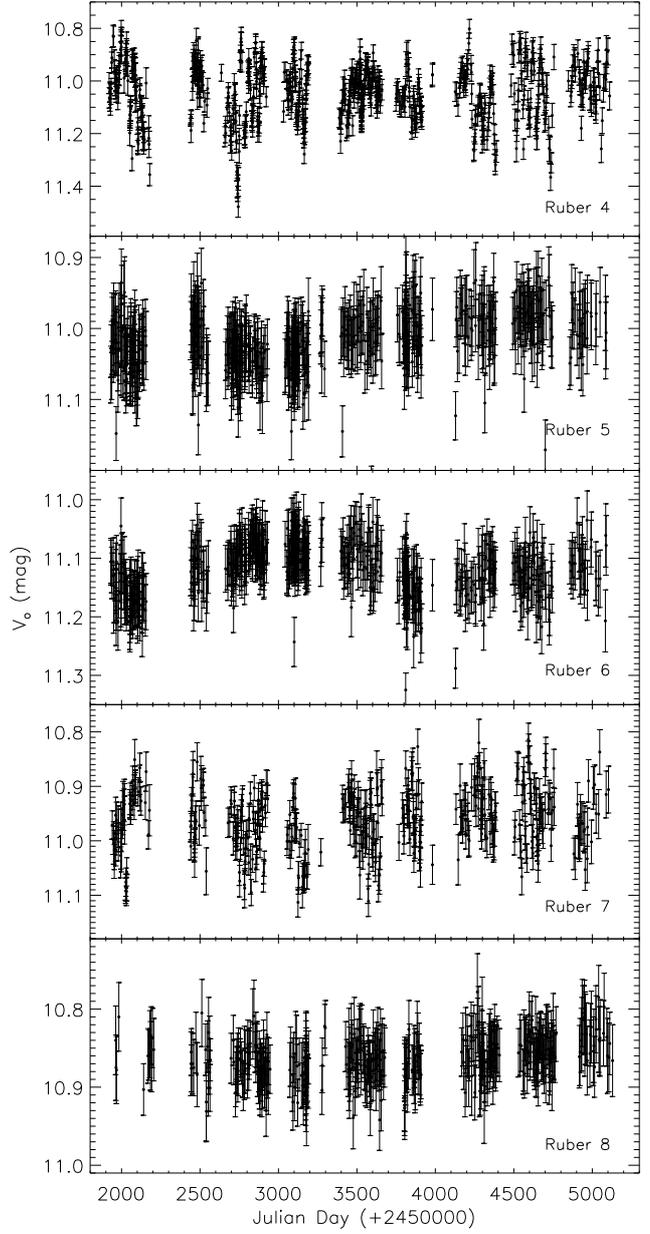}
\caption{ASAS $V$-band light curves of Ruber\,4 to 8, from top to
  bottom.}
\label{lightcurves}
\end{figure}

\begin{figure}
\centering
\includegraphics[width=0.49\textwidth]{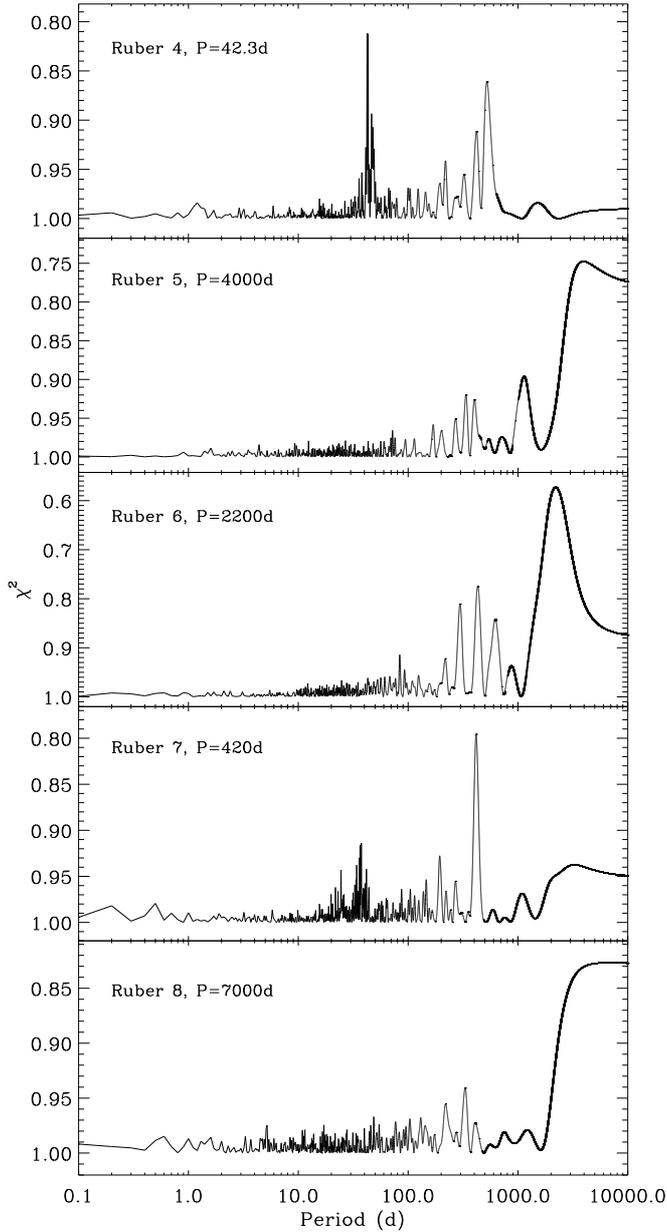}
\caption{Periodograms of the ASAS light curved of Ruber\,4 to 8, from
  top to bottom.}
\label{periodograms}
\end{figure}

\begin{figure}
\centering
\includegraphics[width=0.49\textwidth]{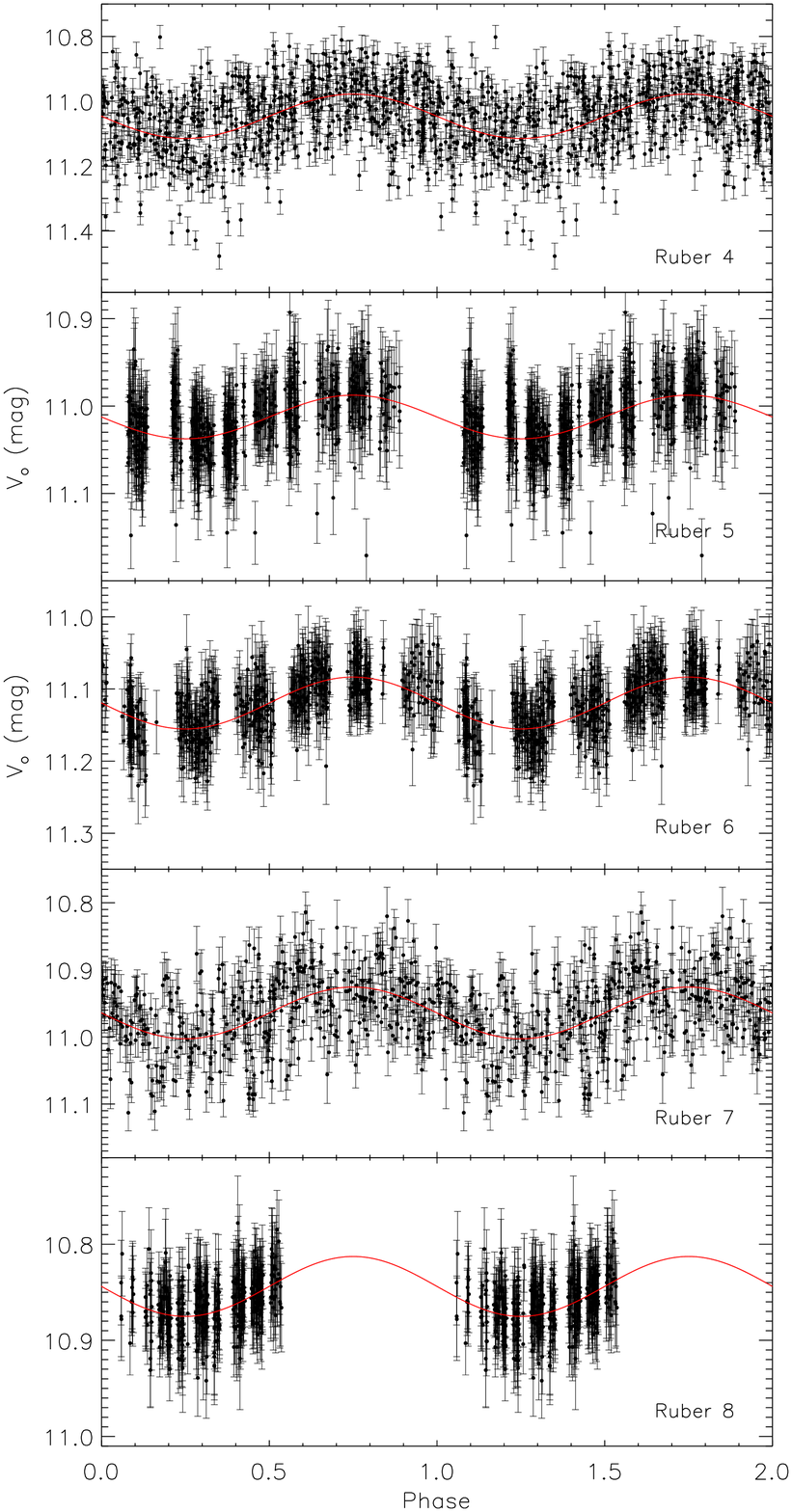}
\caption{ASAS $V$-band light curves as a function of phase of Ruber 4
  to 8, from top to bottom. Continue (red) lines show the
  corresponding symmetric sinusoidal model light curve fit. Light
  curves of Ruber\,5 and 8 are highly tentative.}
\label{lightcurvesphase}
\end{figure}

Another piece of information that may help in classifying the five
Ruber stars in Table~\ref{table.unknown} is their photometric
variability, which is often detected in red giant stars and in
magnetically-active field M dwarfs. In general, active M dwarfs rotate
fast \citep{Stauffer86,Bouvier93,Delfosse98,Rockenfeller06}, while
giants pulsate with long periods
\citep{Stebbins30,Percy98,Koen02,Catelan09}. Periods, if detected, may
also help differentiate the two kinds of objects.

We searched in the All Sky Automated Survey (ASAS) catalogue of
variable stars \citep{Pojmanski02} for data on our objects. One of
them, Ruber\,4 (ASAS~152557--4844.6), was catalogued to have a period
of photometric variability $P$ = 43\,d with an amplitude of 0.30\,mag
in the $V$ band. Tabulated variability type was ``MISC/SR'' (``mostly
semi-regular'' with time scales of variation between 10 and 200\,d).

To do our own time-series analysis, we looked for the original
light-curve data set of the five Ruber targets at the ASAS
webpage\footnote{\tt http://www.astrouw.edu.pl/asas/} as in
\cite{Caballero10a}. The light curves, which contain approximately
between 400 and 700 data points after discarding low-quality
measurements, span from February 2001 to October 2009 and are
displayed in Fig.~\ref{lightcurves}. By eye, most stars display
amplitudes of variability between two and three times the uncertainty
of the photometry. In particular, for Ruber\,4, which is the most
variable star in our small sample, the mean error bar and standard
deviation of the light curve are 40 and 116\,mmag, respectively. The
photometric variability (Tycho-2, 2MASS, ASAS data are not coincident
in time) cannot explain the abnormal observed $V_T-K_{\rm s}$ colours
with respect to the reduced proper motions.

We performed the same variability analysis as in
\cite{Jimenez-Esteban06b}, using a sinusoidal light curve as a first
approximation of the real one. The least mean square method
\citep{Stellingwerf78} was applied for a range of periods from 0.1\,d
to 10,000\,d, and the quality of the fit was determined with a
normalised $\chi^2$ test, weighting each observation with the inverse
of the square of the observational error.

The periodograms for Ruber\,4 to 8 are shown in
Fig.~\ref{periodograms}. As illustrated by the first panel, apart from
confirming the \cite{Pojmanski02} period of Ruber\,4 within
uncertainties ($P$\,=\,42.3$\pm$0.1\,d), we report on the discovery of
a long secondary period (LSP) at around 520\,d. A similar period
($P$\,$\sim$\,420\,d) was found for Ruber\,7, while Ruber\,6 displayed
a period longer than half the total time coverage of 3150\,--\,3200\,d
(cf. bottom of Table~\ref{table.unknown}). For Ruber\,5 and Ruber\,8,
two extremely long periods were found ($P$\,$\sim$\,4000 and
$\sim$\,7000\,d, respectively). Since they are much more longer than
the time coverage of the monitoring, they should be considered highly
tentative so need to be confirmed. All periodograms show a similar
pattern around 300\,--\,400\,d, which is due to the frequency of the
ASAS observations, with blocks of data separated by about a year.

The presence of LSPs, as in the case of Ruber\,4, is very common among
pulsating red giant stars. At least one third of the semi-regular
variables in the solar vicinity are in fact multiperiodic, with an LSP
roughly one order of magnitude longer than the primary period of
pulsation \citep{Wood04,Percy04}. In spite of that, LSP is the only
type of large-amplitude stellar variability that remains completely
unexplained. \cite{Wood04} discuss several mechanisms, but find no
clear explanation for any, and speculate that asymptotic giant branch
stars with LSP may be the precursors of asymmetric planetary
nebulae. More recent studies \citep{Soszynski07a,Soszynski07b} have pointed
towards a binary origin. This last explanation is consistent with the
location of Ruber\,4 in the reduced proper motion diagram, where other
multiple stellar system are found.

Another peculiar object is Ruber\,6, which presents an extremely long
period ($P$\,$\sim$\,2200\,d) with low-amplitude variability
($\sim$0.07\,mag), and relatively blue near-infrared colour ($H-K_{\rm
  s}$\,$\sim$\,0.35\,mag). We did not find any other star with a
longer period reported in the bibliography. Two kinds of red giant
stars could present such long periods: Mira-like variables
\citep{Engels83,Jimenez-Esteban06b} and semi-regular variables
(e.g. \citealt{Glass07}, and references therein). In the first case,
stars present very red near-infrared colours ($H-K_{\rm
  s}$\,$>$\,3\,mag) and a large amplitude of variability ($>$1\,mag);
in the second case, the stars have bluer colours and present lower
amplitudes. Ruber\,6 may be a semi-regular variable star with an
extremely long period and very low-amplitude variability.

\subsection{Spectroscopy}
\label{spectroscopicdata}

The best way to ascertain the nature of the Ruber\,4 to 8 sources is
with spectroscopic data. Consequently, we used another VO tool,
VOSED\footnote{\tt http://sdc.cab.inta-csic.es/vosed}
\citep{Gutierrez08}, to look for spectra in the VO archives. Developed
by the Spanish Virtual Observatory, VOSED allows the user to gather
spectroscopic information available throughout the VO. Unfortunately,
no spectroscopic data were found.

To complement our VO analysis, we collected low-resolution optical
spectra for Ruber\,7 and 8, TYC~6238--480--1 (the discarded low-proper
motion giant in Section~\ref{targetselection}) and eleven comparison
stars with a ground telescope. On 2011 Mar 20, we used the Calar Alto
Faint Object Spectrograph\footnote{\tt
  http://www.caha.es/alises/cafos/cafos.html} (CAFOS) at the 2.2\,m
telescope on the Calar Alto Observatory, Almer\'{\i}a, Spain, with the
grism Green--100 and the SITe\,1d detector of 24\,$\mu$m pixels. The
resulting resolution was only R\,$\sim$\,1600, but the wavelength
coverage was very wide, from 4900 to 7800\,{\AA} without vignetting.
Needed exposure times were in the range of 100\,--\,200\,s. No other
Ruber stars could be observed because of their low declination for
Calar Alto.

\begin{figure*}
\centering
\includegraphics[width=0.49\textwidth]{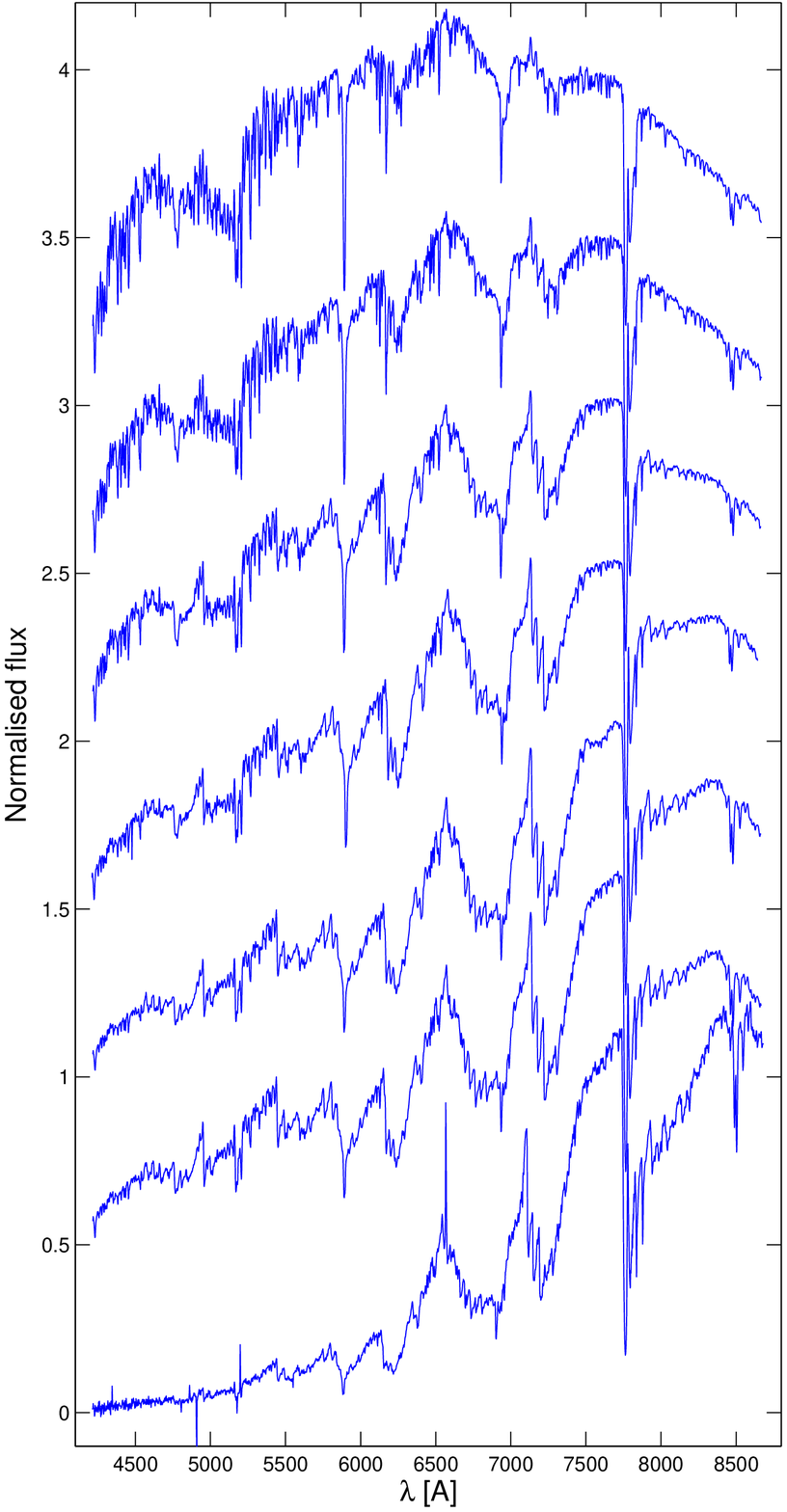}
\includegraphics[width=0.49\textwidth]{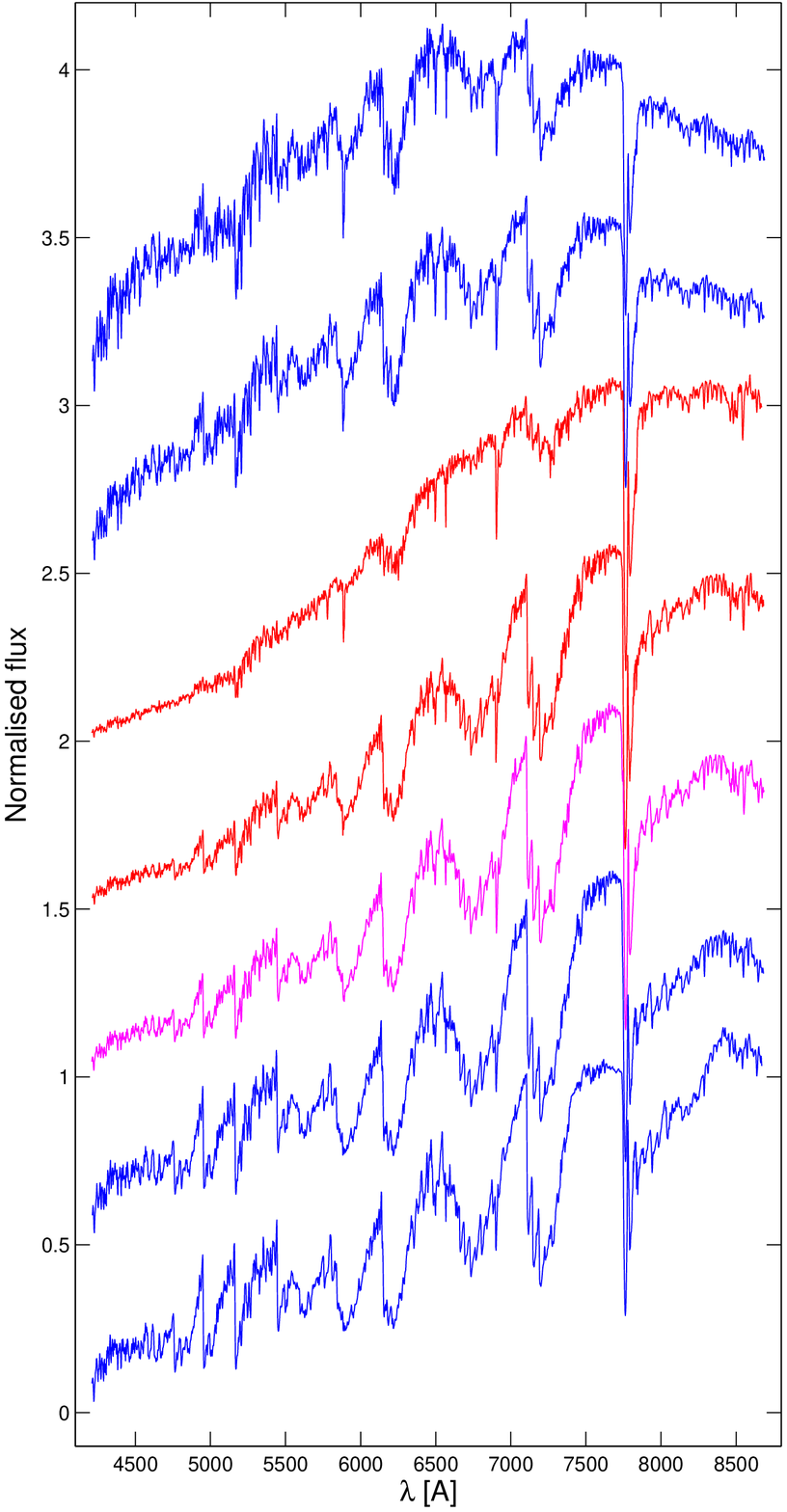}
\caption{CAFOS spectra of Ruber 7 and 8, TYC~6238--480--1, and eleven
  comparison stars.
{\em Left panel} (dwarfs): from top to bottom, 
\object{BD+39~2801} (K5V),
\object{HD~147379~A} (K7V),
\object{GJ~1170} (M1V),
\object{HD~95735} (M2V),
\object{GJ~4040} (M3V),
\object{GJ~687} (M3V), and
\object{GJ~1101} (M3.5Ve).
All the spectral types are from \cite{Hawley96}.
{\em Right panel} (giants): from top to bottom, 
\object{HD~98500} (M0III; \citealt{Moore50}),
\object{HD~122132} (M2III; \citealt{Moore50}),
Ruber\,8 [in red],
Ruber\,7 [in red],
\object{TYC~6238--480--1} [in magenta],
\object{BW~CVn} (M1III; \citealt{Upgren60}), and
\object{BZ~CVn} (M3III; \citealt{Schild73}).}
\label{spectra}
\end{figure*}

The eleven comparison stars were seven nearby late-type dwarfs and
four M-type giants. We reduced, extracted, and corrected all the
spectra for instrumental response (with the standard star Feige~34)
using standard tasks within the IRAF environment.  The fourteen
spectra are shown in Fig.~\ref{spectra}.

From the visual inspection of the pseudocontinuum of the CAFOS
spectra, we concluded that both Ruber\,7 and~8 are in the giant
class. Therefore, we estimated their spectral types from the
comparison with the four other observed giants. Unfortunately, their
spectral type determinations do not seem to be fully reliable: the
spectra of BW~CVn and BZ~CVn are identical, but the reported spectral
types are different (M1III vs. M3III). Something similar happens to
HD~98500 and HD~122132. After consulting additional M giant standards
in \cite{Danks94}, we gave the M2III spectral type, with one dex
uncertainty, to both Ruber\,7 and TYC~6238$-$480$-$1. However, the
spectrum of Ruber\,8 deserved further attention.

The pseudocontinuum of Ruber\,8 reasonably fits that of other M2
giants. However, its absorption bands and (alkali) lines are far less
marked, so we gave it the ``wk'' code of spectral peculiarity (from
``weak lines''). This weakness of bands and lines may be explained by
a very low metallicity. Metal-poor stars are thought to be the
survivors of the earliest generations of stars. Their study helps to
put constraints, for example, on the chemical history of the Milky Way
\citep{Hollek11,Cayrel06}. Bright low-metallicity stars are very rare,
so if the metal-poor nature of Ruber\,8 is confirmed, it would become
an excellent target for detailed follow-up studies. Until now, we have
failed to obtain a higher resolution spectrum of Ruber\,8, from which
we would measure radial velocities, determine galactocentric space
velocities $UVW$, and assign membership for the Galactic kinematic
components (i.e. thin/thick disc, halo).

\section{Summary and final remarks}

We identified red high proper-motion objects in the Tycho-2 and 2MASS
catalogues using Virtual Observatory tools, in the same way as
\cite{Jimenez-Esteban11} did with blue objects. After discarding six
sources with erroneous proper motions and eight close binaries, we got
a sample of 59 objects with proper motions larger than
50\,mas\,yr$^{-1}$ and red colours $V_T-K_{\rm
  s}$\,$>$\,5.5--6.0\,mag. Of them, 54 were known to be M-type giants
and dwarfs. The other five, namely Ruber\,4 to~8, are studied in this
work for the first time.

We collected and analysed all available data of the Ruber objects.
From SED theoretical fits, we estimated temperatures that correspond
to late-type M stars and surface gravities midway between dwarfs and
giants (subject to uncertainties in the used models), but from
near-infrared colours, we concluded that all of them are likely red
giants with high tangential velocities.

The analysis of their ASAS light curves yielded interesting
results. We established reliable periods for three of the five Ruber
objects. In the case of Ruber\,4, we also found a secondary period
that is roughly ten times longer than the primary one. Ruber\,6
presents an extremely long period (P~$\sim$~2200\,d), although with a
large error due to the short time coverage. Ruber\,5 and 8 present
even longer periods, but they are uncertain.

With the help of low-resolution spectra obtained with CAFOS on Calar
Alto, we determined the spectral type of Ruber\,7 and 8 at M2III: and
M2III:\,wk, respectively. The spectrum of Ruber\,8 displays low marked
absorption bands and lines, probably due to a metal-poor nature.

Because of their brightness, the five Ruber objects can serve as
useful targets for detailed studies of old-population giants.


\begin{acknowledgements}
We are in debt to D. Montes for supervising the spectroscopic
observations on Calar Alto, to A. Klutsch for trying to get additional
spectra for Ruber objects, and to C. Rodrigo for clarifying which
models are actually available in VOSA. This research made use of VOSA
and VOSED, developed by the Spanish Virtual Observatory through grants
AyA2008-02156 and RI031675, and of Aladin and SIMBAD developed at the
Centre de Donn\'ees astronomiques de Strasbourg, France. Financial
support was provided by the Spanish Ministerio de Ciencia e
Innovaci\'on, Universidad Complutense de Madrid and Comunidad
Aut\'onoma de Madrid under grants CSD2006-00070 (under the
Consolider-Ingenio 2010 Programme {\em First Science with the Gran
  Telescopio Canarias}), AyA2008-00695, 
AyA2008-06423-C03-03, 
and SP2009/ESP-1496. 
This research made use of the NASA/IPAC Infrared Science Archive,
which is operated by the Jet Propulsion Laboratory, California
Institute of Technology, under contract with the National Aeronautics
and Space Administration. Based on observations collected at the
Centro Astron\'omico Hispano Alem\'an (CAHA) at Calar Alto, operated
jointly by the Max-Planck Institut f\"ur Astronomie and the Instituto
de Astrof\'{\i}sica de Andaluc\'{\i}a (CSIC).
\end{acknowledgements}

\bibliographystyle{aa} 
\bibliography{/pcdisk/muller/fran/RESEARCH/bibliography/references}

%

\longtabL{1}{
\begin{landscape}
\begin{longtable}{lcccccccclll}
  \caption{\label{table.known.giants} Already-known bright red high proper-motion objects: giants.}\\
    \hline\hline
    \noalign{\smallskip}
TYC            & $\alpha$      & $\delta$      & $V_T$             &$J$      &$H$      &$K_{\rm s}$& $\mu$          & $\pi$     & Spectral      & Name           & Ref.$^{a}$    \\
               & (J2000)       & (J2000)       & [mag]             & [mag]   & [mag]   & [mag]    & [mas\,a$^{-1}$] & [mas]    & type          &                &               \\
    \noalign{\smallskip} 
    \hline
    \noalign{\smallskip}
    \endfirsthead
    \caption{continued.}\\
    \hline\hline
    \noalign{\smallskip}
TYC            & $\alpha$      & $\delta$      & $V_T$             &$J$      &$H$      &$K_{\rm s}$& $\mu$          & $\pi$     & Spectral      & Name           & Ref.$^{a}$    \\
               & (J2000)       & (J2000)       & [mag]             & [mag]   & [mag]   & [mag]    & [mas\,a$^{-1}$] & [mas]    & type          &                &               \\
    \noalign{\smallskip}
    \hline
    \noalign{\smallskip}
    \endhead
    \hline
    \endfoot
5846--131--1   & 00 21 46.27  & --20 03 28.9  &  5.926$\pm$0.009  & ~~0.5:  & --0.5:  & --0.8:  & ~~64.6$\pm$1.0  & ~~3.7$\pm$0.5  & M5--6SIIe     & \object{T Cet}         & 1  \\ 
4681--469--1   & 01 14 14.83  & --02 10 46.2  &  8.694$\pm$0.014  & ~~3.0:  & ~~2.0:  & ~~1.7:  & ~~60.4$\pm$1.1  &  ...           & M4III         & \object{AN Cet}        & 2  \\ 
2308--2170--1  & 01 58 03.77  &  +31 08 03.8  &  7.157$\pm$0.010  & ~~1.5:  & ~~0.4:  & ~~0.2:  & ~~62.8$\pm$0.8  & ~~5.1$\pm$0.5  & M5III         & \object{AA Tri}        & 2  \\ 
5278--1494--1  & 02 00 26.82  & --08 31 25.9  &  5.656$\pm$0.009  & ~~0.8:  & --0.2:  & --0.5:  & ~~88.7$\pm$0.8  & ~~6.8$\pm$0.3  & M3III         & \object{AR Cet}        & 3  \\ 
630--507--1    & 02 02 38.63  &  +07 40 36.5  &  9.96$\pm$0.03    & ~~1.4:  & ~~0.4:  & ~~0.0:  & ~~54.3$\pm$1.7  & ...            & M7III         & \object{BD+06 319}     & 4  \\ 
4693--1144--1  & 02 19 20.79  & --02 58 39.5  &  6.654$\pm$0.010  & --0.7:  & --1.6:  & --2.2:  &  238.6$\pm$0.7  & 10.9$\pm$1.2   & M2--7IIIe+DA  & \object{Mira Cet AB}   & 5  \\ 
641--24--1     & 02 53 46.22  &  +09 20 08.9  &  6.816$\pm$0.011  & ~~1.7:  & ~~0.6:  & ~~0.5:  & ~~61.5$\pm$0.8  & ~~5.8$\pm$0.6  & M6III         & \object{EG Cet}        & 6  \\ 
8054--1103--1  & 02 53 52.77  & --49 53 22.7  &  8.253$\pm$0.013  & ~~0.7:  & --0.1:  & --0.6:  &  133$\pm$2      & ~~4.8$\pm$1.0  & M7IIIe        & \object{R Hor}         & 7  \\ 
5338--890--1   & 05 11 22.87  & --11 50 56.7  &  5.869$\pm$0.010  & --0.2:  & --1.1:  & --1.4:  & ~~64.5$\pm$0.9  & ~~6.7$\pm$0.4  & M6III         & \object{RX Lep}        & 8  \\ 
6530--1428--1  & 06 53 00.30  & --26 57 27.5  &  6.586$\pm$0.009  & ~~1.6:  & ~~0.6:  & ~~0.3:  & ~~66.1$\pm$1.3  & ~~3.6$\pm$0.5  & M6III         & \object{KX CMa}        & 9  \\ 
7642--1461--1  & 07 13 32.32  & --44 38 23.1  &  4.767$\pm$0.009  & --1.1:  & --1.9:  & --2.3:  &  342.9$\pm$1.2  & 15.6$\pm$1.0   & M5IIIe        & \object{L$_2$ Pup}     & 10 \\ 
8921--184--1   & 07 19 14.65  & --67 11 15.2  &  8.174$\pm$0.012  & ~~3.4:  & ~~2.5:  & ~~2.1:  & ~~50.8$\pm$1.0  & ~~2.4$\pm$0.5  & M4III         & \object{WW Vol}        & 11 \\ 
2451--1111--1  & 07 20 45.69  &  +31 27 32.9  &  8.81$\pm$0.02    & ~~3.6:  & ~~2.8:  & ~~2.4:  & ~~64.2$\pm$1.3  & ...            & M5:III        & \object{BD+31 1540}    & 12 \\ 
9381--22--1    & 07 55 21.37  & --76 29 52.3  &  8.786$\pm$0.014  & ~~3.4:  & ~~2.4:  & ~~2.0:  & ~~54.0$\pm$1.3  & ...            & M5III         & \object{HD 66424}      & 2  \\ 
6015--2511--1  & 08 39 53.54  & --17 18 10.7  &  6.846$\pm$0.011  & ~~0.6:  & --0.3:  & --0.6:  &  151.7$\pm$1.3  & ~~6.4$\pm$0.4  & M6III         & \object{AK Hya}        & 13 \\ 
818--577--1    & 09 08 26.54  &  +13 13 13.6  &  9.271$\pm$0.016  & ~~1.4:  & ~~0.4:  & ~~0.1:  & ~~60.7$\pm$1.2  & ~~4.1$\pm$0.9  & M6III         & \object{CW Cnc}        & 14 \\ 
8968--918--1   & 10 32 38.42  & --66 10 54.3  &  8.971$\pm$0.015  & ~~2.7:  & ~~1.8:  & ~~1.4:  & ~~50.8$\pm$1.4  & ...            & M6III:        & \object{HD 91600}      & 2  \\ 
5495--421--1   & 10 37 51.80  & --12 01 15.3  &  8.997$\pm$0.017  & ~~1.9:  & ~~1.0:  & ~~0.7:  & ~~54$\pm$2      & ...            & M6III         & \object{FF Hya}        & 15 \\ 
6653--720--1   & 11 14 37.78  & --26 04 30.8  &  7.739$\pm$0.011  & ~~2.7:  & ~~1.8:  & ~~1.5:  & ~~65.3$\pm$1.4  & ...            & M4III         & \object{HY Hya}        & 16 \\ 
7202--37--1    & 11 18 50.10  & --30 28 25.4  & 10.05$\pm$0.03    & ~~2.9:  & ~~2.0:  & ~~1.6:  & ~~60.7$\pm$1.7  & ...            & M8III         & \object{V444 Hya}      & 17 \\ 
8972--291--1   & 11 37 34.06  & --60 54 11.6  &  6.764$\pm$0.010  & ~~1.2:  & ~~0.3:  & ~~0.0:  & ~~67.3$\pm$1.3  & ~~6.1$\pm$0.5  & M4III         & \object{V913 Cen}      & 18 \\ 
292--330--1    & 12 30 21.01  &  +04 24 59.1  &  8.328$\pm$0.015  & ~~0.5:  & --0.4:  & --0.7:  & ~~53.2$\pm$1.2  & ~~5.5$\pm$0.7  & M7III         & \object{BK Vir}        & 19 \\ 
5540--333--1   & 13 11 20.83  & --10 30 50.8  &  7.640$\pm$0.013  & ~~2.8:  & ~~1.9:  & ~~1.6:  & ~~55.7$\pm$1.2  & ...            & Mb...         & \object{V339 Vir}      & 20 \\ 
899--580--1    & 13 37 52.93  &  +13 26 48.4  &  8.614$\pm$0.015  & ~~2.5:  & ~~1.6:  & ~~1.3:  & ~~58.9$\pm$0.9  & ~~4.1$\pm$0.7  & M6III         & \object{DH Boo}        & 21 \\ 
8269--1422--1  & 13 39 59.81  & --49 56 59.8  &  6.044$\pm$0.009  & ~~0.5:  & --0.6:  & --0.8:  &  101.0$\pm$0.9  & ~~6.4$\pm$0.3  & M6III         & \object{V744 Cen}      & 22 \\ 
6728--19--1    & 13 49 02.00  & --28 22 03.5  &  7.788$\pm$0.012  & --1.7:  & --2.7:  & --3.2:  & ~~78.3$\pm$1.3  & ~~9.6$\pm$1.1  & M8IIIe        & \object{W Hya}         & 23 \\ 
7287--1891--1  & 13 49 26.72  & --34 27 02.8  &  4.416$\pm$0.009  & --0.6:  & --1.5:  & --1.8:  & ~~73.9$\pm$1.1  & 17.8$\pm$0.2   & M4.5III       & \object{$g$ Cen}       & 24 \\ 
1467--355--1   & 13 53 55.19  &  +17 16 50.8  &  9.58$\pm$0.03    & ~~4.1:  & ~~3.1:  & ~~2.8:  & ~~50.7$\pm$1.2  & ...            & M5III         & \object{XZ Boo}        & 25 \\ 
6724--764--1   & 13 55 19.31  & --26 25 57.4  &  8.018$\pm$0.012  & ~~2.5:  & ~~1.6:  & ~~1.3:  & ~~83.0$\pm$1.2  & ~~0.3$\pm$0.6  & M5III         & \object{V349 Hya}      & 26 \\ 
9427--2476--1  & 14 05 19.88  & --76 47 48.3  &  5.791$\pm$0.009  & --0.7:  & --1.7:  & --2.1:  & ~~94.4$\pm$0.7  & ~~8.8$\pm$0.5  & M6.5III       & \object{$\theta$ Aps}  & 27 \\ 
1478--509--1   & 14 46 05.95  &  +15 07 54.4  &  6.021$\pm$0.010  & ~~0.5:  & --0.4:  & --0.7:  & ~~87.7$\pm$0.9  & ~~4.0$\pm$0.4  & M5IIIab       & \object{EK Boo}        & 28 \\ 
5594--455--1   & 15 19 21.81  & --09 08 47.5  &  7.223$\pm$0.011  & ~~2.2:  & ~~1.4:  & ~~1.1:  & ~~51.8$\pm$1.1  & ~~2.8$\pm$1.0  & M4III         & \object{FZ Lib}        & 29 \\ 
2578--824--1   & 15 58 30.77  &  +36 01 19.7  &  7.606$\pm$0.010  & ~~2.5:  & ~~1.6:  & ~~1.3:  & ~~51.6$\pm$1.4  & ~~3.0$\pm$0.4  & M7III         & \object{RS CrB}        & 30 \\ 
3491--136--1   & 16 02 39.17  &  +47 14 25.3  &  6.624$\pm$0.010  & --0.1:  & --0.9:  & --1.3:  & ~~94.8$\pm$1.1  & ~~7.3$\pm$0.4  & M6IIIe        & \object{X Her}         & 31 \\ 
4190--653--1   & 16 35 00.72  &  +60 28 05.3  &  7.498$\pm$0.011  & ~~2.7:  & ~~1.7:  & ~~1.5:  & ~~58.0$\pm$1.3  & ~~2.9$\pm$0.5  & M4IIIe        & \object{TX Dra}        & 32 \\ 
8717--558--1   & 16 42 19.86  & --55 23 03.7  &  8.828$\pm$0.015  & ~~3.6:  & ~~2.6:  & ~~2.2:  & ~~71$\pm$2      & ...            & M3--4II:      & \object{HD 150184}     & 33 \\ 
5087--261--1   & 17 53 03.32  & --02 34 45.6  &  7.710$\pm$0.012  & ~~1.7:  & ~~0.8:  & ~~0.5:  & ~~56.9$\pm$0.9  & ~~2.7$\pm$0.7  & M6III         & \object{V533 Oph}      & 34 \\ 
1029--3054--1  & 18 39 57.12  &  +09 58 21.9  &  8.056$\pm$0.012  & ~~2.0:  & ~~1.2:  & ~~0.9:  & ~~54$\pm$2      & ...            & M6IIIe        & \object{HD 172450}     & 35 \\ 
3131--2155--1  & 18 55 20.10  &  +43 56 45.9  &  4.355$\pm$0.009  & --0.7:  & --1.6:  & --1.8:  & ~~83.7$\pm$0.7  & 10.94$\pm$0.12 & M5III         & \object{R Lyr}         & 36 \\ 
461--458--1    & 18 57 18.34  &  +06 41 53.4  &  9.57$\pm$0.02    & ~~3.6:  & ~~2.5:  & ~~2.2:  & ~~65$\pm$2      & ...            & M7III         & \object{V840 Aql}      & 37 \\ 
1040--241--1   & 19 06 22.25  &  +08 13 48.0  &  8.225$\pm$0.014  & ~~0.7:  & --0.4:  & --0.8:  & ~~72.2$\pm$1.6  & ~~2.4$\pm$0.9  & M7IIIev       & \object{R Aql}         & 38 \\ 
8782--316--1   & 19 43 13.64  & --56 15 37.0  &  7.914$\pm$0.011  & ~~2.1:  & ~~1.3:  & ~~0.9:  & ~~54.5$\pm$1.3  & ~~5.1$\pm$0.6  & M5--M6III     & \object{V341 Tel}      & 39 \\ 
7942--935--1   & 19 58 42.87  & --41 50 57.9  &  8.272$\pm$0.014  & ~~2.8:  & ~~2.0:  & ~~1.6:  & ~~58.0$\pm$1.1  & ~~2$\pm$2      & M4IIIe        & \object{RU Sgr}        & 40 \\ 
6913--836--1   & 20 06 55.25  & --27 13 29.8  &  8.179$\pm$0.012  & --0.1:  & --1.1:  & --1.5:  & ~~51.3$\pm$1.6  & ~~5.1$\pm$0.6  & M8III         & \object{V1943 Sgr}     & 41 \\ 
5743--34--1    & 20 10 07.02  & --10 32 12.5  &  8.92$\pm$0.02    & ~~3.5:  & ~~2.5:  & ~~2.2:  & ~~53.5$\pm$1.3  &  ...           & M3III         & \object{HD 191429}     & 42 \\ 
1637--2033--1  & 20 37 54.73  &  +18 16 06.9  &  6.362$\pm$0.010  & ~~0.3:  & --0.8:  & --1.0:  & ~~72.1$\pm$1.0  & ~~8.6$\pm$0.5  & M6III         & \object{EU Del}        & 43 \\ 
4460--2400--1  & 21 09 31.78  &  +68 29 27.2  &  8.338$\pm$0.012  & --0.5:  & --1.3:  & --1.8:  & ~~63.7$\pm$1.2  & ~~5.3$\pm$0.9  & M7IIIe        & \object{T Cep}         & 44 \\ 
3591--3422--1  & 21 36 02.50  &  +45 22 28.5  &  6.179$\pm$0.010  & --0.2:  & --1.1:  & --1.4:  & ~~64.5$\pm$1.0  & ~~5.7$\pm$0.4  & M4III         & \object{W Cyg}         & 45 \\ 
7495--579--1   & 22 04 42.40  & --35 41 46.9  &  8.038$\pm$0.012  & ~~2.7:  & ~~1.8:  & ~~1.6:  & ~~56.8$\pm$1.3  & ~~3.4$\pm$0.7  & M5III         & \object{VW PsA}        & 46 \\ 
8827--195--1   & 22 57 05.86  & --57 24 04.2  &  4.564$\pm$0.010  & ~~1.3:  & ~~0.4:  & ~~0.0:  & ~~87.9$\pm$1.4  & ~~4.5$\pm$0.5  & M8III         & \object{DM Tuc}        & 47 \\ 
\footnote{$^{a}$ Representative references --
     (1) \cite{Sanner78}; 
     (2) \cite{Kwok97}; 
     (3) \cite{Adams26}; 
     (4) \cite{Winters03}; 
     (5) \cite{Campbell1899}; 
     (6) \cite{Olnon86}; 
     (7) \cite{Fleming1892}; 
     (8) \cite{Libert08}; 
     (9) \cite{Eggen70}; 
     (10) \cite{Bedding02}; 
     (11) \cite{Koen02}; 
     (12) \cite{Espin1892}; 
     (13) \cite{Eggen73}; 
     (14) \cite{Sivagnanam04}; 
     (15) \cite{Hoffmeister33}; 
     (16) \cite{Stokes71}; 
     (17) \cite{Hashimoto94}; 
     (18) \cite{Blanco55}; 
     (19) \cite{LeBouquin09}; 
     (20) \cite{Strohmeier66}; 
     (21) \cite{Guglielmo97}; 
     (22) \cite{Cool05}; 
     (23) \cite{McGee77}; 
     (24) \cite{Ruban09}; 
     (25) \cite{Jura92b}; 
     (26) \cite{Platais03}; 
     (27) \cite{Pickering1898}; 
     (28) \cite{Konstantinova-Antova10}; 
     (29) \cite{Lebzelter01}; 
     (30) \cite{Mennesson05}; 
     (31) \cite{Matthews11}; 
     (32) \cite{Solowieff28}; 
     (33) \cite{Houk75}; 
     (34) \cite{Feast72}; 
     (35) \cite{Vyssotsky46}; 
     (36) \cite{Tscherny41}; 
     (37) \cite{Chen01}; 
     (38) \cite{Cotton10}; 
     (39) \cite{Adelman01}; 
     (40) \cite{Whitelock00}; 
     (41) \cite{Begemann97}; 
     (42) \cite{Matsuura99}; 
     (43) \cite{Percy89}; 
     (44) \cite{Sanner77}; 
     (45) \cite{Curtiss04}; 
     (46) \cite{Ramirez97}; 
     (47) \cite{Hacking85}; 
}
\end{longtable}
\end{landscape}
}

\begin{landscape}
\begin{table}
  \caption[]{Already-known bright red high proper-motion objects: dwarfs.}
  \label{table.known.dwarfs}
  \begin{center}
  \begin{tabular}{l cc cccccc lll}
    \hline
    \hline
    \noalign{\smallskip}
TYC           & $\alpha$    & $\delta$     & $V_T$           & $J$             & $H$             & $K_{\rm s}$       & $\mu$          & $\pi$  	   & Spectral  & Name           & Ref.$^{c}$\\
              & (J2000)     & (J2000)      & [mag]           & [mag]           & [mag]           & [mag]            & [mas\,a$^{-1}$] & [mas]     	   & type      &                & \\
    \noalign{\smallskip}                                                                                                                                                                                   
    \hline                                                                                                                                                                                                 
    \noalign{\smallskip}                                                                                                                                                                                   
5913--1376--1 & 05 06 49.48 & --21 35 03.7 & 11.71$\pm$0.12  & 7.00$\pm$0.02    & 6.39$\pm$0.02    & 6.11$\pm$0.02    & ~~~~53$\pm$4    & 100:$^{a}$  & M3.5V+M4.5V & \object{BD--21 1074 BC} & 1,2,3 \\ 
2944--1956--1 & 07 10 01.84 &  +38 31 46.2 & 11.8$\pm$0.2    & 6.73$\pm$0.03    & 6.15$\pm$0.05    & 5.85$\pm$0.02    & 1043$\pm$4      & ~~159$\pm$3 & M4.5Ve SB2  & \object{QY Aur AB}      & 4,5,6 \\ 
183--2190--1  & 07 44 40.15 &  +03 33 09.0 & 11.35$\pm$0.12  & 6.58$\pm$0.02    & 6.00$\pm$0.04    & 5.698$\pm$0.017  & ~~567$\pm$2     & ~~168$\pm$2 & M4.0V       & \object{YZ CMi AB}      & 7,8,9 \\ 
2003--139--1  & 13 31 46.61 &  +29 16 36.7 & 12.3$\pm$0.2    & 7.56$\pm$0.02    & 7.00$\pm$0.02    & 6.72$\pm$0.02    & ~~272$\pm$3     & ~~80:$^{b}$ & M4.0V+M:    & \object{DG CVn AB}      & 10,11,12 \\ 
  \noalign{\smallskip}
  \hline
  \end{tabular}
  \end{center}
  \begin{list}{}{}
     \item[$^{a}$] Photometric parallax from \cite{Reid95b}. 
     \item[$^{b}$] Photometric parallax estimated from \cite{Beuzit04}. 
     \item[$^{c}$] Representative references --
     (1) \cite{Stephenson86};
     (2) \cite{Reid04};
     (3) \cite{Jao03};
     (4) \cite{Ross39};
     (5) \cite{Pettersen75};
     (6) \cite{Tomkin86};
     (7) \cite{Ross37};
     (8) \cite{Raassen07};
     (9) \cite{Kowalski10};
     (10) \cite{Robb99};
     (11) \cite{Beuzit04};
     (12) \cite{Reiners10}.
  \end{list}
\end{table}
\end{landscape}

\begin{table*}
  \caption[]{New bright red high proper-motion objects$^{a}$.}
  \label{table.unknown}
  \begin{center}
  \begin{tabular}{l c cccccc}
    \hline
    \hline
    \noalign{\smallskip}
    Ruber              		& $\overline{\lambda}$	& 4                     & 5                     & 6                     & 7                     & 8            \\
                  		& [$\mu$m] 		&                       & (BD--19~4369)         &                       &                       &              \\
    \noalign{\smallskip}                                                                                                                        
    \hline                                                                                                                                      
    \noalign{\smallskip}                                                                                                                        
    TYC                                      &  	& 8299--2512--1         & 6211--498--1          & 6211--472--1          & 5648--545--1          & 5113--1090--1 \\
    IRAS                                     &  	& 15223--4834           & R16234--1931          & 16249--1919           & 17099--0950           & R18397--0123  \\
    $\alpha$ (J2000)                         &  	& 15:25:56.87           & 16:26:21.17           & 16:27:50.71            & 17:12:40.72           & 18:42:20.54 \\
    $\delta$ (J2000)                        &   	& --48:44:35.1          & --19:38:43.9          & --19:26:07.1           & --09:54:12.1          & --01:20:15.2 \\
    $l$ (deg)                               &   	& 327.42                & 356.85                & 357.26                &  12.09                &  30.73                \\
    $b$ (deg)                               &   	&  +6.64                & +20.03                & +19.90                & +16.73                &  +1.47                \\
    $\mu_\alpha \cos{\delta}$ [mas\,a$^{-1}$]&   	& +61$\pm$3             & --58$\pm$3            & --117$\pm$3~           & --31$\pm$3~~          & --33$\pm$2~~ \\
    $\mu_\delta$ [mas\,a$^{-1}$]             &   	& --8$\pm$3             & --2$\pm$3             & --63$\pm$4             & --40$\pm$3~~          & --47$\pm$2~ \\
    $H_{V_{T}}$ [mag]                        &   	& 10.04$\pm$0.18       & 10.08$\pm$0.18         & 11.89$\pm$0.13        &  9.60$\pm$0.19        &  9.64$\pm$0.15 \\
    \noalign{\smallskip}                                                                                                                       
    \hline                                                                                                                                      
    \noalign{\smallskip}                                                                                                                       
    \multicolumn{7}{c}{Photometric data} \\
    \hline                                                                                                                                      
    \noalign{\smallskip}                                                                                                                       
    Tycho-2 $B_T$ [mag]             	& 0.42    	&  12.8$\pm$0.3         & 14.3$\pm$0.6          & 13.9$\pm$0.5          & 12.9$\pm$0.3             & 12.4$\pm$0.2 \\
    Tycho-2 $V_T$ [mag]             	& 0.53    	& 11.07$\pm$0.09        & 11.26$\pm$0.10        & 11.27$\pm$0.10        & 11.08$\pm$0.09           & 10.82$\pm$0.08 \\
    UCAC3 $r$ [mag]               	& 0.63    	& 10.22$\pm$0.02        & 10.52$\pm$0.06        & 10.62:                & 10.30$\pm$0.02           & 10.035$\pm$0.013  \\
    DENIS $i$ [mag]               	& 0.80    	&  8.84$\pm$0.04        &  9.25$\pm$0.04        & ...                   &  8.37$\pm$0.02           &  8.80$\pm$0.05 \\
    2MASS $J$ [mag]               	& 1.23    	&  6.52$\pm$0.02        &  6.82$\pm$0.02        &  6.61$\pm$0.02        &  6.192$\pm$0.018         &  6.184$\pm$0.019 \\
    2MASS $H$ [mag]               	& 1.66    	&  5.56$\pm$0.02        &  5.94$\pm$0.04        &  5.66$\pm$0.03        &  5.22$\pm$0.04           &  5.15$\pm$0.03  \\
    2MASS $K_s$ [mag]             	& 2.16    	& 5.191$\pm$0.017       & 5.625$\pm$0.017       &  5.31$\pm$0.03        &  4.81$\pm$0.02           &  4.81$\pm$0.02 \\
    {\em WISE} $W1$ [mag]               & 3.35    	&  5.05$\pm$0.06        &  5.49$\pm$0.05        &  5.16$\pm$0.06        &  4.76$\pm$0.06           &  4.69$\pm$0.06 \\
    {\em WISE} $W2$ [mag]               & 4.60    	&  4.88$\pm$0.04        &  5.24$\pm$0.03        &  4.70$\pm$0.03        &  4.32$\pm$0.03           &  4.28$\pm$0.04 \\
    GLIMPSE $5.8$ [mag]           	& 5.76    	& ...                   & ...                   &  ...                  &  ...                     &  4.59$\pm$0.03 \\
    GLIMPSE $8.0$ [mag]           	& 7.96    	& ...                   & ...                   &  ...                  &  ...                     &  4.58$\pm$0.02 \\
    {\em MSX} $A$ [Jy]                  & 8.28    	& ...                   & ...                   &  ...                  &  ...                     &  0.85$\pm$0.04 \\
    {\em AKARI} $S09$ [Jy]              & 9.22    	& 0.581$\pm$0.009       & 0.437$\pm$0.014       &  ...                  &  0.81$\pm$0.02           &  0.82$\pm$0.04 \\
    {\em WISE} $W3$ [mag]               & 11.56   	&  4.92$\pm$0.02        &  5.36$\pm$0.02        &  5.08$\pm$0.02        &  4.62$\pm$0.02           &  4.63$\pm$0.02 \\
    {\em IRAS} $12$ [Jy]                & 11.60   	&  0.49$\pm$0.04        & ...                   &  0.38$\pm$0.05        &  0.77$\pm$0.18           &  ...    \\
    {\em AKARI} $S18$ [Jy]              & 19.81   	&  0.15$\pm$0.03        & ...                   &  ...                  &  0.23$\pm$0.07           &  0.208$\pm$0.003 \\
    {\em WISE} $W4$ [mag]               & 22.09   	&  4.83$\pm$0.03        &  5.31$\pm$0.03        &  5.00$\pm$0.03        &  4.53$\pm$0.03           &  4.61$\pm$0.03 \\
    \noalign{\smallskip}
    \hline                                                                                                                                      
    \noalign{\smallskip}
    \multicolumn{7}{c}{Estimated parameters$^{b}$} \\
    \hline                                                                                                                                      
    \noalign{\smallskip}                                                                                                                       
    $T_{\rm eff}$ [K]          &       & 3400$\pm$100  	& 2900$\pm$100  & 3000$\pm$100   & 3100$\pm$100   & 3200$\pm$100 \\ 
    $\log{g}$ [cgs]           &       & 3.5$\pm$0.5     & 3.5$\pm$0.5   & 3.5$\pm$0.5    & 3.5$\pm$0.5    & 4.0$\pm$0.5  \\
    $P$ [d]                   &       & 42.3$\pm$0.3    & (4000)        & 2200$\pm$800   & 420$\pm$30     & (7000)       \\
    Amplitude [mmag]            &      & 126$\pm$9       & (50)          &  70$\pm$40      &  80$\pm$50    & (60)        \\
    Sp. type                  &       & ...             & ...           & ...            & M2III:         & M2III:\,wk \\
  \hline
  \end{tabular}
  \end{center}
  \begin{list}{}{}
     \item $^{a}$ Tycho-2 and {\em IRAS} identification, equatorial
       coordinates from 2MASS, Galactic coordinates and proper motions
       from Tycho-2; magnitudes and fluxes from catalogues listed in
       the text; $T_{\rm eff}$ and $\log{g}$ from our SED fits; period
       and amplitude in the $V$ band from our time-series analysis;
       and spectral type from our CAFOS observations.
     \item $^{b}$ Tentative periods and amplitudes in parenthesis.
  \end{list}
\end{table*}

\end{document}